\documentclass[twoside,12pt,reqno]{amsart}
\usepackage[margin=1 in]{geometry}
\usepackage{amsmath,amsthm, amsfonts,amssymb}
\usepackage{braket}
\usepackage{color,soul,mathtools}
\usepackage[dvipsnames]{xcolor}
\usepackage[all]{xy}
\usepackage{graphicx}
\usepackage{indentfirst}
\usepackage{bm}
\usepackage{mathrsfs}
\usepackage{latexsym}
\usepackage{hyperref}
\usepackage{cleveref}
\usepackage{bbm}

\title{Kraus-Like Decompositions}

\author{Jonathan Boretsky}
\address{\textbf{J. Boretsky:} Department of Mathematics, Harvard University, Cambridge, MA 02138}

\author{Robert Lin}
\address{\textbf{R. Lin:} Department of Mathematics, Harvard University, Cambridge, MA 02138; Department of Chemistry and Chemical Biology, Harvard University, Cambridge, MA 02138}

\newtheorem{definition}{Definition}
\newtheorem{theorem}[definition]{Theorem}
\newtheorem{proposition}[definition]{Proposition}
\newtheorem{cor}[definition]{Corollary}
\newtheorem{example}[definition]{Example}
\newtheorem{lemma}[definition]{Lemma}

\numberwithin{definition}{section}

\begin{document}

\begin{abstract}
We introduce a new decomposition of quantum channels acting on group algebras, which we term Kraus-like (operator) decompositions. We motivate this decomposition with a general nonexistence result for Kraus operator decompositions in this setting. Given a length function which is a class function on a finite group, we construct a corresponding Kraus-like decomposition. We prove that this Kraus-like decomposition is \textit{convex} (meaning its coefficients are nonnegative and satisfy a sum rule) if and only if the length is conditionally negative definite. For a general finite group, we prove a stability condition which shows that the existence of a convex Kraus-like decomposition for all $t>0$ small enough  necessarily implies existence for all time $t>0$. Using the stability condition, we show that for a general finite group, conditional negativity of the length function is equivalent to a set of semidefinite linear constraints on the length function. Our result implies that in the group algebra setting, a semigroup $P_t$ induced by a length function which is a class function is a quantum channel for all $t\geq 0$ if and only if it possesses a convex Kraus-like decomposition for all $t>0$.

\end{abstract}

\maketitle

\section{Introduction}

The notion of a completely positive, trace-preserving map (CPTP map), also called a \textit{quantum channel}, is important in a variety of areas, including quantum information theory and the study of various inequalities for operator algebras. Quantum channels are often studied in particular mathematical settings, such as full matrix algebras or other particular kinds of $C^*$ algebras. The setting of full matrix algebras is of particular importance in quantum information theory via the study of finite-dimensional density matrices and their properties (with respect to norms, entropies, etc.). In such a situation, quantum channels are characterized by the well-known Kraus operator decomposition theorem \cite{Sti}\cite{Cho}\cite{MikeIke}.  %{\color{brown}(NB: try to include discussion of general Stinespring dilation.)} %one has a natural setting for quantum information theory through the study of finite-dimensional density matrices and their properties (with respect to norms, entropies, etc.), and completely positive, trace-preserving maps (which define a \textit{quantum channel}) are characterized by the well-known Kraus decomposition theorem.

In this article, we specialize to the particular case of group algebras where the underlying group is finite\footnote{The group algebra setting has previously been studied in quantum computation in Kitaev's quantum double model for finite groups \cite{Kitaev} (see  \cite{Cui} for recent results in this direction).}. Quantum channels in the group algebra setting possess similar desirable properties as those in the full matrix algebra setting, and thus are of independent interest.  For example, the interpolation theory of Uhlmann \cite{Uhl} extends the monotonicity of the relative entropy under identity-preserving completely positive maps from the usual matrix algebra setting to arbitrary $*$-algebras, which contains group algebras as a special case.  We prove a nonexistence result for a Kraus operator decomposition of the quantum channel in terms of Kraus operators lying within the group algebra. This motivates us to introduce operators which allow one to decompose certain %kinds of completely positive maps on the group algebra
quantum channels acting on the group algebra. The decomposition we introduce involves a sum over operators which come with real-valued coefficients.  Much as the work of Choi \cite{Cho} shows that the existence of a Kraus operator decomposition in the matrix algebra case implies that the corresponding linear mapping is a quantum channel, we will obtain a result in the group algebra case showing that, for particular naturally arising operator semigroups $P_t$, if a decomposition of $P_t$ in terms of the operators we introduced satisfies a positivity condition on the coefficients, then the elements of the semigroup are quantum channels for all $t\geq 0$. Hence, we term these operators \textit{Kraus-like operators} and the corresponding decompositions \textit{Kraus-like operator decompositions}. We further define \textit{convex} Kraus-like operator decompositions to be those in which the coefficients in the sum decomposition are nonnegative and satisfy a particular sum rule, which we will discuss explicitly later.%In particular, we will study Kraus-like operator decompositions of a semigroup $P_t$ of particular interest.

% For quantum channels induced by certain kinds of length functions defined on the group, we show that that there is a decomposition in terms of multipliers which are diagonal in the group algebra basis with coefficients analogous to  probabilities. The appearance of probability-like coefficients is particular to the choice of multipliers. While motivated by Kraus operator decompositions, our decomposition is quite different from the usual Kraus operator decomposition of a quantum channel. Thus, we term the operator decomposition a \textit{Kraus-like operator decomposition}. When the coefficients are probability-like, i.e. the coefficients arising in the decomposition are all nonnegative and satisfy a sum rule, we term the operator decomposition a \textit{convex Kraus-like operator decomposition}, or simply \textit{convex Kraus-like decomposition}.

By the work of \cite{Sch}, it is known that the imposition of a conditionally negative-definite length on the finite group underlying the group algebra results in the induced semigroup $P_t$  (specifically, take the semigroup of operators $P_t$ defined by $P_t \lambda(g) = e^{-t l(g)} \lambda(g)$ where $\lambda(g)$ is a multiplier and $l$ is a scalar-valued function on the group, and extend by linearity to all of $\mathcal{L}G$) being a quantum channel for all $t\geq 0$. By restricting to length functions which are class functions, and passing to the Kraus-like operator decomposition, we show that the condition of conditional negative-definiteness can be efficiently verified by checking a number of semidefinite linear constraints on the length function. We show that the latter is a necessary and sufficient condition for conditional negative-definiteness, and thus for the semigroup to be a quantum channel for all $t\geq 0$.

The choice of multipliers in our framework is canonical, as we induce the multipliers by the characters of the finite group upon which the group algebra is built. This enables us to use representation theory of finite groups to achieve our results. %Some structure must be built into the group in order to make the induced semigroup a quantum channel.

%are some physical interpretations in terms of probabilities one may make of certain kinds of completely positive maps defined on a group algebra. This probabilistic interpretation may \Jon{lead to more concrete connections from group algebras to physics.}be helpful to making group algebras ``useful" for more concrete connections to physics. 

%The motivation for us to introduce these Kraus-like operators is the following result: for certain semigroups acting on a group algebra, Kraus decompositions of the semigroup in terms of left-multiplication Kraus operators cannot exist. This motivates us to introduce Kraus-like decompositions, which are decompositions of a semigroup into a linear combination of left-multiplication operators \Jon{We should probably give a similar one sentence run down of Kraus decompositions, perhaps in the first paragraph}.  We show that convex Kraus like decompositions that are induced by the irreducible characters of a finite group are equivalent to conditionally negative length such that the length is a class function. Furthermore, we show that for a general finite group, the convexity of the character-induced Kraus-like decomposition is equivalent to a set of semidefinite \Jon{linear} constraints on the length function, thus providing a \textit{probabilistic} interpretation of conditional negativity \Jon{The connection here is not obvious to me.}. 

In terms of our proof method, for a general finite group $G$, in order to obtain our semi-definite linear constraints, we prove a stability condition which shows that if the semigroup associated with a length  has a convex Kraus-like operator decomposition for all $t>0$ small enough, then it has a convex Kraus-like operator decomposition for all time $t>0$. Thus, to obtain global positivity of the coefficients, it suffices to check positivity near $t=0^{+}$, which reduces to a bound on the derivatives. This derivative bound is what yields the semi-definite linear constraints.

%Since, by a theorem of Schoenberg, conditional negativity of a length guarantees complete positivity of the corresponding semigroup, our connection between convex Kraus-like operator decompositions and conditional negativity allows us to establish several equivalent formulations of complete positivity of a semigroup induced by a length function that is a class function.

%In terms of broader context, since our work pertains to neither full matrix algebras, nor to the kinds of von Neumann algebras as might arise in quantum field theory. Hence, a priori, the physical meaning is less clear, .

\section{Definitions}

The main objects we are working with in this paper are the left regular representation of a finite group $G$, and a semigroup acting on the elements in the left regular representation. Later in this section, we will find it useful to restrict to 
semigroups which are induced by conditionally negative-definite length functions on $G$, in a way to be precisely defined. %Then, we will motivate somewhat different constraints on length functions, which we will show are equivalent to conditional negativity under the assumption that the length function is a class function. 
%These notions will all be defined in this section. These different constraints allow one to obtain a more physical \textit{probabilistic} picture of conditional negativity.

\begin{definition}[Left Regular Representation]
Given a group $G$, let $\mathcal{F}G$ be the vector space of complex-valued functions on $G$. We denote by $\lambda$ the \textbf{left regular representation} of $G$, which acts on  $\mathcal{F}G$ by: $(\lambda(g) f)(h) = f( hg^{-1}) $ for $g\in G$ and $f \in \mathcal{F}G$. Denote the $\mathbb{C}-$linear span of $\{\lambda(g)\}_{g\in G}$ by $\mathcal{L}G$.
\end{definition}
\noindent As it is a property we will use repeatedly, we emphasize that by definition of a representation, for each $g,h$ in $G$, we have the equality $\lambda(g)\lambda(h)=\lambda(gh)$ of operators on $\mathcal{F}G$. 

Recall the standard inner product on $\mathcal{F}G$, given by $\left<f,h\right>_{\mathcal{F}G}=\sum_{g\in G}\overline{f(g)}h(g)$ for $f,h\in\mathcal{F}G$. The space $\mathcal{L}G$ also comes with a natural inner product. Let $\delta_e$ be the function on $G$ defined by $\delta_e(g) = \delta_{g=e}$. Then, we define $\langle x, y \rangle_{\mathcal{L}G} = \langle \delta_e, x^* y \,\delta_e\rangle_{\mathcal{F}G}$, where the $*$ operation is defined on $\mathcal{L}G$ antilinearly with $\sum_i\alpha_i \lambda(g_i) \mapsto \sum_i\bar{\alpha_i} \lambda(g_i^{-1})$. It is straightforward to verify that this is in fact an inner product and thus, $(\mathcal{L}G,\braket{\cdot,\cdot}_{\mathcal{L}G})$ is a Hilbert space. 

To define a semigroup on $\mathcal{L}G$, one needs the notion of a length function on $G$. Following \cite{JPPP}, we restrict ourselves to conditionally negative-definite lengths.
\begin{definition} A \textbf{length function} $l:G\rightarrow \mathbb{R}$ on a group $G$ with identity $e$ is a function which satisfies $l(e)=0$, $l(g)=l(g^{-1})$ and $l(g)\geq 0$ for all $g$ in $G$.
\end{definition}
If $l(g)>0$ for all $g\neq e$, then we will call $l$ a \textit{strict} length function. 

\begin{definition}[\cite{For}]
A length function $l:G\rightarrow\mathbb{R}$ is said to be \textbf{conditionally negative-definite} if for any $\alpha_1,\cdots \alpha_n\in\mathbb{C}$ satisfying $\sum_{i=1}^n\alpha_i=0$ and any $g_1,\cdots,g_n\in G$, one has 
    \begin{equation*}
        \sum_{i,j=1}^n \alpha_i\overline{\alpha_j}l(g_i^{-1}g_j)\leq 0.
    \end{equation*}
    
    \end{definition}
%    Note that by symmetry in $i$ and $j$, the above expression is always real, and so it may be compared to 0.

    An important additional assumption we adopt in this work is the assumption that all length functions are class functions, meaning they are constant on conjugacy classes. Symbolically, this means $l(g)=l(h^{-1}gh)$ for any $g,h$ in $G$. This assumption will be greatly exploited in our results and we will derive new characterizations of these conditionally negative-definite class function lengths.
    
    We consider the semigroup $P_t$ of operators on $\mathcal{L}G$ induced by a length function $l(\cdot)$ which is given on generators of $\mathcal{L}G$ by the action
    \begin{equation}
    \label{semigroup}
    P_t \lambda(g) = e^{-t\, l(g)} \lambda(g),
\end{equation}
\noindent and extended linearly. Observe that indeed, $P_0$ acts by the identity and $P_{t_1}P_{t_2}=P_{t_1+t_2}$. 

Our goal is to characterize $P_t$ from various perspectives. Our new perspectives on $P_t$ will shed light on a known characterization of $P_t$ \cite{JPPP} presented in the continuation of this section. 

% as a completely positive map when $l$ is conditionally negative, which follows from Schoenberg's theorem. 

% We define:
    %Let us first present one characterization which isBefore we proceed further, we  recall some structural facts about conditionally negative lengths, which will be useful later on. We first need the notion of a positive definite kernel.
    \begin{definition}
    A Hermitian function $K:G\times G\rightarrow \mathbb{C}$ is said to be a \textbf{positive definite kernel} if for any $\alpha_1,\cdots,\alpha_n\in \mathbb{C}$ and any $g_1,\cdots,g_n\in G$, we have 
    
    \begin{equation*}
        \sum_{i,j=1}^n\alpha_i\overline{\alpha_j}K(g_i,g_j)\geq 0.
    \end{equation*}
\end{definition}
 %   Once again, note that by symmetry in $i$ and $j$, the above expression is always real, and so it may be compared to 0. 

%Having established this terminology, we are now able to state Schoenberg's theorem, which will play a key role in our analysis of abelian groups. 
Schoenberg's theorem provides a characterization of positive definite kernels in terms of conditionally negative-definite lengths:
\begin{theorem}[Schoenberg's Theorem \cite{For}]
    Let $G$ be a group. A function $l:G\rightarrow \mathbb{R}$ satisfying $l(g)=l(g^{-1})$ for all $g\in G$ is conditionally negative definite if and only if the following conditions hold: 
    
    \begin{enumerate}
        \item $l(e)=0$, for $e$ the identity of $G$, and
        \item The function $G\times G\rightarrow \mathbb{C}$ defined by $(g,h)\mapsto e^{-t\,l(gh^{-1})}$ is positive definite.
    \end{enumerate}
    
\end{theorem}

In \Cref{condnegequivcompos}, we recall an equivalent characterization of conditionally negative-definite lengths in terms of the notion of complete positivity, which is important in quantum physics and quantum information theory. Recall that a linear map is \textbf{positive} if it maps positive elements to positive elements. %an important notion is that of a \textbf{completely positive} map between Hilbert spaces, which is relevant in our context. 
Following \cite{Nesh}, 
\begin{definition}
Let $\mathcal{A}$ and $\mathcal{B}$ be $C^*$ algebras, and $\theta: \mathcal{A}\rightarrow \mathcal{B}$ be a linear map. The map $\theta$ is called \textbf{completely positive} if 
\begin{equation}
    \theta \otimes \text{id}: \mathcal{A} \otimes \text{Mat}_{n}(\mathbb{C}) \rightarrow \mathcal{B} \otimes \text{Mat}_{n}(\mathbb{C})
\end{equation}
is positive for any $n \in \mathbb{N}$. %Define $\phi_n: M_n(\mathcal{A})\rightarrow M_n(\mathcal{B})$ by $\phi_n((a_{i,j})) = (\phi(a_{i,j}))$. We call $\phi$ \textbf{n-positive} if $\phi_n$ is positive, and we call $\phi$ \textbf{completely positive} if $\phi$ is n-positive for all $n$.
\end{definition}
A consequence of Schoenberg's theorem is that the semigroup  $P_t$ 
is completely positive if and only if $l(\cdot)$ is conditionally negative-definite (stated, but not proved, in \cite{JPPP}). To be complete, we provide a proof:
\begin{proposition}\label{condnegequivcompos}
 $P_t$ is completely positive if and only if $l$ is conditionally negative-definite.
\end{proposition}
\begin{proof}
%\section{Proof}
($\Leftarrow$) By Schoenberg's theorem, if $l$ is conditionally negative-definite, then $\left(e^{-t\,l(g^{-1} h)}\right)_{g,h\in G}$ is a positive semi-definite matrix. From appendix A of \cite{Nesh}, $P_t: \mathcal{L}G \rightarrow \mathcal{L} G$ is completely positive if and only if
\begin{equation}
	\sum_{i,j=1}^{n} b_i^* P_t(a_i^* a_j)b_j \geq 0
\end{equation}
for any $n\in \mathbb{N}$, $a_1, \ldots, a_n, b_1, \ldots,b_n \in  \mathcal{L} G$. We show that the latter holds. 

Fix $n$. Take $a_i = \sum_{g\in G} a_i(g) \lambda(g)$. Then 
\begin{equation}
	P_t (a_i^* a_j) = \sum_{g, h \in G} a_i(g)^* a_j(h) e^{-t\, l(g^{-1}h)}\lambda(g^{-1}h).
\end{equation}
So we want to show that
\begin{equation}
	S := \sum_{i,j=1}^{n}	\sum_{g, h \in G} b_i^*a_i(g)^* a_j(h)  e^{-t\, l(g^{-1}h)}\lambda(g^{-1}h) b_j
\end{equation}
is non-negative. We note that this can be rearranged by setting $v_g = \sum_{i=1}^{n} a_i(g) \lambda(g) b_i$. So the sum becomes
\begin{equation}
	\sum_{g,h\in G} v_g^* e^{-t\, l(g^{-1} h)} v_h.
\end{equation}
Note that since $\left(e^{-t\,l(g^{-1} h)}\right)_{g,h\in G}$ is positive semi-definite, we have that 
\begin{equation}
	e^{-t l(g^{-1} h)} = \sum_{x\in G} r_x(t) \left(w_x(t)^{*} w_x(t)\right)_{g,h}
\end{equation}
for some matrices $\{w_x(t)|x\in G\}$, and nonnegative numbers $\{r_x(t)|x\in G\}$. 

Thus,
\begin{align}
S &=\sum_{g,h\in G} v_g^* \sum_{x\in G} r_x(t) \left(w_x(t)^{*} w_x(t)\right)_{g,h} v_h \\
&= \sum_{x\in G} r_x(t) \sum_{g,h\in G} v_g^* \left(w_x(t)^{*} w_x(t)\right)_{g,h}  v_h \\
&= \sum_{x\in G} r_x(t) \sum_{g,h\in G} v_g^* \sum_{m\in G} \left(w_x(t)^{*}\right)_{g,m} \left(w_x(t)\right)_{m,h}  v_h
\end{align}
Set
\begin{equation}
	q_{x,m}(t) = \sum_{g\in G} \left(w_x(t)\right)_{m,g} v_g
\end{equation}
then
\begin{equation}
	S = \sum_x r_x(t) \sum_{m\in G} q_{x,m}^{*}(t) q_{x,m}(t).
\end{equation}
Thus, $S$ is positive semi-definite, as desired.

($\Rightarrow$) Conversely, suppose $P_t$ is completely positive. Then, with notation as above, for all $n\in\mathbb{N}$ and $a_1,\cdots, a_n,b_1,\cdots, b_n\in\mathcal{L}G$, 

\begin{equation}\label{posimpcondneg}
    S\coloneqq \sum_{i,j=1}^nb_i^*P_t(a_i^*a_j)b_j=\sum_{g,h\in G}v_g^*e^{-t\ell(g^{-1}h)}v_h\geq 0, 
\end{equation}
where $v_g\coloneqq \sum_{i=1}^na_i(g)\lambda(g)b_i$. Since this holds for any choice of $n$, $a_i$ and $b_i$, we can fix $n=|G|$. Order the elements of $G$ so that $G=\{g_1,\cdots,g_n\}$. Choose $b_i=\lambda(g_i^{-1})$ and choose $a_i$ such that $a_i(g_j)=c_j\delta_{ij}$ for some $c_j\in \mathbb{C}$. Then, for each $j=1,\cdots n$, $v_{g_j}=c_j\lambda(e)$. This reduces \cref{posimpcondneg} to 

\begin{equation*}
    \left(\sum_{i,j=1}^n \overline{c_i}e^{-t\ell(g_{i}^{-1}g_{j})}c_j\right)\lambda(e)\geq 0,
\end{equation*}
for any choice of $c_1,\cdots c_n\in \mathbb{C}$. Now we claim that if $A\lambda(e)\geq 0$ with $A\in \mathbb{C}$, then $A\geq 0$. This follows since $\lambda(e)$ acts as the identity, and so its spectrum is just $1$, and so the spectrum of $A\lambda(e)$ is just $A$. Thus, for any choice of $c\in\mathbb{C}^n$, we have 

\begin{equation*}
    \sum_{i,j=1}^n \overline{c_i}e^{-t\ell(g_{i}^{-1}g_{j})}c_j\geq 0.
\end{equation*}

By definition, this means that the matrix $\{e^{-t\ell(g_{i}^{-1}g_{j})}\}_{i,j=1}^n$ is positive semi-definite and so, by Schoenberg's theorem, $\ell$ is a conditionally negative-definite length. 
\end{proof}

\begin{cor}
\label{quantumchannel}
$P_t$ is a quantum channel.
\end{cor}
\begin{proof}
Since $P_t$ is also trace-preserving, it follows by definition that $P_t$ is a quantum channel \cite{MikeIke}.
\end{proof}

\section{Kraus-Like Operator Decompositions}

In quantum information theory, \textit{Kraus operators} are used in sum representations of quantum channels which describe the dynamics of the density matrix of a system \cite{MikeIke}. In fact, in the matrix case, quantum channels can be characterized by the existence of a Kraus operator decomposition. Equivalently, one may describe a quantum channel as the result of tracing out a subsystem from a unitary operator acting on a composite system. Conversely, one may always ``lift" a quantum channel on a density matrix to a corresponding unitary operator on a larger system, a process known as Stinespring dilation \cite{Sti}. 

One of the main questions which drives this work is whether the usual intuition for quantum channels on density matrices extends to those on group algebras. Namely, does one get Kraus operators? And what do they look like?

What we find is that the Kraus operators, if they exist, are certainly not generally elements of the group algebra. This is in direct contrast to the case of quantum channels acting on density matrices, where the Kraus operators are themselves matrices.

This nonexistence result motivates us to look for alternate decompositions of the quantum channel, in the spirit of the Kraus operator decomposition. The main idea is to relax the action of Kraus operators as $E\cdot E^{\dagger}$ to some linear operator $\sigma \cdot$, where $\sigma$ acts on $\mathcal{L}G$. We call our new $\sigma$'s \textit{Kraus-like} operators.

For a suitable choice of $\sigma$'s, we can obtain an explicit condition on the decomposition of a semigroup $P_t$ induced by a length function $l$ which is also a class function, which determines whether or not $P_t$ is a quantum channel.

\subsection{An algebraic obstruction result on Kraus operator decompositions}
 Consider the semigroup $P_t$, induced by a length $l$,  which acts on  the Hilbert space $\mathcal{H}=\mathcal{L}G$. We recall that this is given as the linear extension of:
\begin{equation}
    P_t \lambda(g) = e^{-t l(g)} \lambda(g).
\end{equation}  

%Since, for any $g\in G$, $\lambda(g)$ acts on $\mathcal{L}G$, we have an action of $\mathcal{L}G$ on itself and so $\mathcal{L}G$ can be identified as a subspace of $\mathcal{B}({\mathcal{H}})$. It follows that $P_t$ is an element of $\mathcal{B}({\mathcal{H}})$. \Jon{Do we need all this? Isn't $P_t$ an element of $\mathcal{B}(H)$ just by consequence of being a linear map?}

A Kraus decomposition of $P_t$ is a decomposition of $P_t$ given by
\begin{equation}
    P_t(x) = \sum_{i} E_i x E_i^{\dagger},
\end{equation}
where the elements $E_i$ satisfy $\sum_i E_i^{\dagger} E_i \leq I$ \cite{MikeIke}. The $E_i$ are called Kraus operators.  For simplicity, we focus only on the case where $\sum_{i} E_{i}^{\dagger} E_i = I$, corresponding to the case where one can dilate $P_t$ to a unitary in the matrix case \cite{MikeIke}. In the usual matrix algebra setting for Kraus operator decompositions, quantum information theory, $x$ would be a finite-dimensional density matrix and $P_t$ would be a completely positive map from density matrices to density matrices. The elements $E_i$ would be matrices. For the group algebra setting, it is not \textit{a priori} obvious where the $E_i$'s would live, so we will make some natural assumptions.

Since we are working with group algebras, to employ Kraus operator decompositions, some choices must be made as to the proper identification of terms. The natural mapping, extending the setting of matrix algebras to direct sum of matrix algebras, is to take $P_t$ to map $\mathcal{L}G$ into $\mathcal{L}G$. While one could also embed $\mathcal{L}G$ into a matrix algebra by the regular representation, this is fairly unnatural from the point of respecting the symmetry of the group, as different irreducible representations ought not to interact from a physical perspective. %For instance, one must be clear about the domain and range of $P_t$ \Jon{I think this needs to be expanded upon/clarified}.  Additionally, it must be specified what kind of element $x$ is. 

For a Kraus operator decomposition, since we work in $\mathcal{}G$, it is further natural, or at least convenient, to assume that the $E_i$'s all lie in $\mathcal{L}G$. Note that this is not the most general setting, since if we interpret $\mathcal{L}G$ as a vector space of dimension $|G|$, the dimension of $\mathcal{B}(\mathcal{L}G)$ is $|G|^2$, whereas the embedding of $\mathcal{L}G$ inside $\mathcal{B}(\mathcal{L}G)$ only has dimension $|G|$. So we are deliberately choosing to focus on a smaller space of possible Kraus operators. However,  we show that with this simple, and perhaps most natural, choice, we run into issues.

The following proposition shows that a Kraus operator  decomposition as described in the previous paragraph is not the right tool for the job at the hand, in the sense that Kraus operator  decompositions will not be readily available in many semigroups of interest. %For this purpose, we will fix the simplest possible length a finite group $G$ can admit, defined by $|e|=0$ and $|g|=1$ for $g\neq e\in G$.

\begin{proposition}
Any finite group $G$ whose group algebra $\mathcal{L}G$ has a non-zero element of the form $\sum_{e\neq g\in G} a_g\lambda(g)$ in its center will not admit a Kraus operator decomposition for the operator $P_t$ induced by a \textbf{strict} length function $l$ via equation \ref{semigroup}, in terms of Kraus operators lying in $\mathcal{L}G$. 
\end{proposition}
In particular, the hypothesis of this proposition holds for the expression $\sum_{e\neq g\in G}\lambda(g)\in \mathcal{L}G$ when $G$ is any nontrivial finite group. 
\begin{proof}
Let $h$ be an element satisfying the conditions of the proposition. Suppose $P_t$ admits a Kraus decomposition of the form $P_t(x)=\sum_i E_ixE_i^\dagger$, where $E_i \in \mathcal{L}G$. Since $\lambda(e)$ is the identity, it follows that \begin{equation*}
    \lambda(e)=\exp(-l(e)t)\lambda(e)=P_t\lambda(e)=\sum_iE_iE_i^\dagger.
\end{equation*}

Now, consider $P_t(h)$. Since $h$ is in the center, we have
\begin{equation*}
    P_t(h)=\sum_i E_ihE_i^\dagger=\sum_iE_iE_i^\dagger h=\lambda(e)h=h.
\end{equation*}

However, by assumption, $h=\sum_{g\neq e}a_g\lambda(g)$, so
\begin{equation*}
    P_t(h)=\sum_{g\neq e}e^{-l(g)t}a_g\lambda(g)=h = \sum_{g\neq e}a_g\lambda(g).
\end{equation*}
By the basis property, this implies that $(e^{-l(g)t}-1)a_g=0$ for all $g\neq e$ and for all $t\geq 0$. Since for a length function $l$, $l(g)>0$ for all $g\neq e$, this implies that $a_g=0$ for all $g\neq e$. Thus, $h=0$.

% \begin{equation*}
%     P_t(h)=\sum_{g\neq e}e^{-|g|t}a_g\lambda(g)=e^{-t}\sum_{g\neq e}a_g\lambda(g)=e^{-t}h.
% \end{equation*}
%Thus, for all $t\geq 0$, we have $e^{-t}h=h$, which is a contradiction for non-zero $h$.
\end{proof}

%This nonexistence result motivates us to look for a more general decomposition of a semigroup, which \textit{will} exist even when a Kraus decomposition in terms of group algebra elements is not available. We will present a suggestion for a decomposition where, moreover, it will turn out the coefficients of the decomposition satisfy a sum rule. This opens the door to the possibility of a \textbf{probabilistic interpretation} of the coefficients in our decomposition. \Jon{Should we just get rid of this paragraph and move the part about the probabilistic interpretation to the next paragraph? It does feel like an intro sort of thing anyways!}

\subsubsection{Discussion of dilation theorems}

Some remarks must be made with respect to the Stinespring dilation theorem \cite{Sti}, and the associated Choi isomorphism theorem \cite{Cho}. Firstly, the Stinespring dilation theorem still applies in this context of group algebras, but the construction of a dilation is so general as not to yield anything resembling a Kraus operator decomposition. The much sharper construction of Choi applies in the case of a finite-dimensional Hilbert space. When we look at the group algebras, the theorem of Choi is hard to apply directly, for the following reason: Embedding the group algebra $\mathcal{L}G$ into a matrix algebra via the left regular representation is a sparse isometric embedding\footnote{To see that the embedding is isometric, we can first note that in the case of $\mathbb{Z}_N$, one easily sees that all the elements in the left-regular representation (except the identity) have no fixed points, so $\langle \rho(g), \rho(h) \rangle:=\text{tr}(\rho(g) \rho(h)^{\dagger} =\text{tr}(\rho(g h^{-1})=0$ unless $g=h$. For the general finite group case, an element $\rho(g)$ has an element on the diagonal only if $gh = h$ for some $h$. But this implies that $g=e$ since all elements are invertible in a group. So no non-identity elements of the left-regular representation have elements on the diagonal. Thus, the embedding is isometric in general.}, since the former has a basis set of size $|G|$ whereas the latter has a basis set of size $|G|^2$. Thus, our definition of $P_t$ as a completely positive, trace-preserving map on the group algebra $\mathcal{L}G$ does \textit{not} mean that $P_t$ is a completely positive, trace-preserving map on the corresponding \textit{matrix} algebra, because the action of $P_t$ is not even specified for matrices outside of the left regular representation. What one would be looking for instead is some analogue of Choi's isomorphism theorem, which applies to a map which is completely-positive and trace-preserving on a subalgebra of a matrix algebra. Such a kind of \textit{restriction theorem}\footnote{We are inspired by Stein's restriction conjecture in analysis to use this suggestive vocabulary.} would be interesting in its own right. Of course, one would be working with objects which are not really quantum channels, but only behave like quantum channels when applied to a subalgebra of the matrix algebra (as justified by Corollary \ref{quantumchannel}). The corresponding \textit{extension} problem has been considered recently by \cite{Wir} on an extension result for quantum Markov semigroups (completely positive maps which form a semigroup) defined on a subalgebra to a quantum Markov semigroup over the full matrix algebra.

\subsection{Kraus-like Operator Decompositions}

\subsubsection{What is a Kraus-like Operator Decomposition?}

The above nonexistence result motivates us to look for a more general decomposition of a semigroup, which \textit{will} exist even when a Kraus operator decomposition in terms of group algebra elements is not available. We introduce the notion of a Kraus-like operator decomposition, which replaces the Kraus form by a multiplier on the group algebra. Under several equivalent hypotheses on the length function, we will be able to show that the coefficients of the decomposition satisfy a sum rule and are positive, hence admitting a possible probabilistic interpretation. In this respect, our motivation is to establish something analogous to the mixed-unitary quantum channel for our Kraus-like operator decomposition.

%Having established that for many groups, the operator $P_t$ with an arbitrary length function $l$ does not admit a Kraus decomposition in terms of group algebra elements, we set out to establish an alternative, which we call \textit{Kraus-like decompositions}. 

We now present a prototype for what we consider to be a Kraus-like operator decomposition. Consider a decomposition of the operator semigroup $P_t$ into a sum of isometries, rather than taking a sum of $E_i x E_i^{\dagger}$ operators. Let us show that, at least in a particularly simple example, such a decomposition does in fact exist. 
\begin{example}

Let $G$ be an arbitrary group with the length $l(g)=1-\delta_{g=e}$ for $g\in G$. Let $p=\frac{1-e^{-t}}{2}$. Define the isometry

\begin{equation}
    \sigma(\lambda(g))=\begin{cases}
    -\lambda(g), &g\neq e\\
    \lambda(e), &g=e
    \end{cases}
\end{equation}

Then for any $g\in G$, one can easily verify that $P_t(\lambda(g))=(1-p)\lambda(g)+p\sigma(\lambda(g))$. The $\lambda(g)$ span $\mathcal{L}G$ so this proves that $P_t=(1-p)\mathbb{I}+p\sigma$ as operators on $\mathcal{L}G$.
\end{example}
Let us make a few observations about this decomposition. The coefficients of the two isometries in the decomposition are $(1-p)$ and $p$. We note that by the definition of $p$, these are both non-negative numbers for $t\geq 0$. Moreover, they evidently sum to $1$.

This is a basic example, but we already see that there is hope that our decomposition might have a probabilistic interpretation. %This opens the door to a potential interpretation of $p$ as representing a probability. \Jon{Is the following sentence redundant to the pargraph starting ``We now present a prototype"?}
Motivated by this example, we introduce the following notion as an analog to Kraus operator decompositions: 

\begin{definition}
For a group $G$ and operator semigroup $P_t:\mathcal{L}G\rightarrow\mathcal{L}G$, a \textbf{Kraus-like operator decomposition} of $P$ is a decomposition  \begin{equation}
    P_t=\sum_i p_i(t) \sigma_i
\end{equation} where each operator $\sigma_i: \mathcal{L}G\rightarrow \mathcal{L}G$ is diagonal in the basis of left-multipliers $\lambda(g)$, and $p_i(t)$'s are complex-valued functions.

If the $p_i(t)$'s are all nonnegative, and satisfy a sum rule $\sum_i \alpha_i p_i(t) = 1$, for some positive $\alpha_i$'s independent of $t$, then we say that we have a \textbf{convex Kraus-like operator decomposition}. 
\end{definition}  

The natural next question is if this sort of decomposition can be generalized to the semigroups generated by other lengths. We will show that, indeed, it can. In fact, for a specific class of naturally arising $\sigma_i$ operators, we will even be able to describe a condition which classifies precisely which lengths on a given group will yield semigroups with convex Kraus-like operator decompositions. 

\section{Character-Induced Kraus-Like Operator Decompositions}

\subsection{Kraus-like operator decompositions for finite abelian groups}
Our next goal is to consider a general abelian group $G$ and think more broadly about when we have a convex Kraus-like operator decomposition. 

Fix a finite abelian group $G$ of size $n$. It is natural to consider maps which are induced by the characters of $G$. Explicitly, if the characters of $G$ are denoted by $\{\chi_i\}_{i=1}^n$, then we consider the maps which act on generators by $\sigma_i:\lambda(g)\mapsto\chi_i(g)\lambda(g)$ and extend by linearity. Note that since $G$ is abelian, all characters are simply one-dimensional representations. These are, in fact, isometries and the multiplicative structure they inherit from the fact that they are representations will prove useful later. 

Let $l$ be any length on $G$. Note that the characters of a group span the class functions on that group and, in the case of an abelian group where each element is its own conjugacy class, this means that the characters span the complex-valued functions on the group. Thus, we can write $P_t$ as a sum $P_t = \sum_{k=1}^{n} p_k \sigma_k$ for appropriate $p_k$. By applying both sides of the previous equation to $\lambda(g)$ for each $g\in G$ and comparing coefficients, one finds that the $p_k$ must satisfy:
	\begin{equation}
	\sum_{l=1}^{n}\chi_l(g) p_l = \exp(-tl(g)).
	\end{equation}

Recall that part of our goal in all this is to understand if there is an interpretation of the $p_k$ as probabilities. As we now show, this is equivalent to demanding that $f(g)=\exp(-tl(g))$, as a function of $g\in G$, be of positive type. In other words, we must have that $f(gh^{-1})$ is a positive-definite kernel, when considered as a function of both $g$ and $h$.

\begin{proposition}[Bochner-like Theorem]
	Let $G$ be a finite abelian group of size $n$ with characters $\{\chi_l\}_{l=1}^n$, and suppose $f: G\rightarrow \mathbb{C}$, and $p\in\mathbb{C}^n$ are related by $f(g)=\sum_{l=1}^{n} p_l\chi_l(g)$ for all $g\in G$.

	Then $p$ is a probability measure on $G$, that is, $p_i\geq 0$ for all $i$ and $\sum_ip_i=1$, if and only if $f(gh^{-1})$, considered as a function of $g,h\in G$, is a positive definite kernel, and $f(e)=1$. 
\end{proposition}

\begin{proof}
Note that the positive definiteness of $f$ is equivalent to the non-negativity of the following expression, for any choice of $\varphi: G \rightarrow \mathbb{C}$: 
	
\begin{align*}
    &\sum_{g\in G} \sum_{h\in G} f(gh^{-1}) \varphi(g) \varphi(h)^* =\sum_{l=1}^n p_l \sum_{g\in G} \sum_{h\in G} \chi_l(gh^{-1}) \varphi(g) \varphi(h)^*\\=& \sum_{l=1}^n p_l\left(\sum_{g\in G} \chi_l(g)\varphi(g) \right)\left(\sum_{h\in G} \chi_l(h)\varphi(h) \right)^*=\sum_{l=1}^n p_l\left|\sum_{g\in G} \chi_l(g)\varphi(g) \right|^2, 
\end{align*}
	where the second equality follows from the general fact that $\chi(g^{-1})=\chi(g)^*$ and also the fact that any representation of an abelian group is $1$ dimensional, so the characters of such a group are multiplicative. Note that if each $p_k\geq0$, then this expression is surely non-negative for any choice of $\varphi$. On the other hand, if some $p_k$ is not greater than or equal to $0$, then set $\varphi(g)=\chi_k(g)$. By the orthogonality of characters, this will result in the entire expression failing to be greater than or equal to zero. Thus, we conclude that $f$ is positive definite if and only if each of the $p_l\geq 0$ for all $1\leq l\leq n$.  Additionally, the condition that $\sum_{l=0}^{N-1}p(l)=1$ is equivalent to the condition that $f(e)=1$, completing our proof.
	
% 	Thus, 
% 	\begin{align*}
%     &p \textnormal{ is a probability measure on } G \\
%     \iff &\sum_{l=0}^{N-1}p(l)=1 \textnormal{ and } p(l)\geq 0 \textnormal{ for } l=1,\cdots,n\\
%     \iff&f(e)=1  \textnormal{ and } p(l)\geq 0 \textnormal{ for } l=1,\cdots,n\\
%     \iff & f(e)=1 \textnormal{ and } f \textnormal{ is positive definite.}
% \end{align*}
	
\end{proof}

Our goal was to characterize the lengths $l$ on a finite abelian group $G$ for which we obtain a probability measure $p$ in the convex Kraus-like operator decomposition of the operator semigroup $P_t$ induced by $l$. Using Schoenberg's theorem, we can easily obtain such a condition. 

\begin{proposition}
Let $l$ be any length on a finite abelian group $G$ and let $\{\chi_l\}_{l=1}^n$ be the characters of $G$. Suppose the semigroup $P_t$ is defined by $P_t:\lambda(g)\mapsto e^{-t|g|}\lambda(g)$, which extends linearly to all of $\mathcal{L}G$. Let $P_t$ satisfy $P_t=\sum_{k=1}^{n}p_k\sigma_k$, where $\sigma_k(\lambda(g))=\chi_k(g)\lambda(g)$ for all $g\in G$. Then $p$ is a probability measure on $G$ if and only if $l$ is a conditionally negative-definite length.
\end{proposition}
\begin{proof}
This follows by combining the previous proposition with Schoenberg's theorem, which says that $l$ is conditionally negative-definite if and only if $f(gh^{-1})=\exp(-tl(gh^{-1}))$ is a positive-definite kernel in $g,h\in G$ which satisfies $f(e)=1$. 
\end{proof}

Thus, we have established a nice coherent story for finite abelian groups. We have characterized a large class of lengths on these groups which admit a convex Kraus-like operator decomposition. However, this is hardly satisfying. For one thing, it is not clear whether the relationship between a conditionally negative-definite length and a convex Kraus-like decomposition is an accident or has some more fundamental significance. Moreover, it would be valuable to move this analysis beyond abelian groups. Thus, we explore next the notion of Kraus-like operator decompositions for general finite groups. %subject to some mild and reasonable conditions, we will attempt to provide a more general and easily testable condition for when we may obtain a convex Kraus-like decomposition. 

\subsection{Kraus-like decompositions for general finite groups}
 In the more general setting of finite groups, there are two basic questions we need to answer. 
 \begin{enumerate}
     \item  What is the appropriate generalization of the Kraus-like operator decomposition to a general finite group $G$, based on the model we considered for $\mathbb{Z}_N$?
     \item What is the corresponding condition for the coefficients $p_i$ arising in the decomposition to be non-negative, or more specifically, for the $p_i$'s to be a probability distribution (at least up to rescaling)?
 \end{enumerate}
 %The second is, 
 
% \subsubsection{Framework}
 The answer to the question (1) is that we can simply define multipliers induced by characters in the following way:
Since we suppose that $l$ is a class function, $P_t$ acts as a constant multiple of the identity on the left-multipliers $\lambda(g)$ for $g\in C_i$, for each conjugacy class $C_i$. Accordingly,
\begin{equation}
P_t = \oplus_i \left(e^{-t\, l(C_i)} \otimes 1_{\# C_i}\right),
\end{equation}
where $l(C_i)$ is the unique value of the length function on the conjugacy class $C_i$. 

We may also use the irreducible characters $\chi$ of the group $G$, which are themselves class functions, to induce maps with similar direct sum decompositions,  
\begin{equation}
\sigma_{\chi} := \oplus_i \left(\chi(C_i) \otimes 1_{\# C_i}\right).
\end{equation}

From the similar forms of these operators, it seems reasonable to study relationships of the form
\begin{equation}\label{krauslikedecomp}
P_t = \sum_{\chi} p_{\chi} \sigma_{\chi},    
\end{equation}
for complex numbers $p_\chi$. We call this expression a \textit{character-induced Kraus-like} operator decomposition of the semigroup $P_t$.

Let $\{\chi_1,\cdots, \chi_m\}$ be the irreducible characters of $G$. We can quickly determine the coefficients $p_j$. By applying the map $P_t=\sum_j p_{j}\sigma_{\chi_j}$ to $\lambda(g)$ for $g\in C_i$ and setting the two sides equal to each other, one obtains, for each $i$, the equation:

\begin{equation}
\label{matrixeqn}
\sum_j p_j \chi_{ji} = e^{-t\, l(C_i)}
\end{equation}
where $\chi_{ji} = \chi_j(C_i)$. We can consider these as entries of a matrix $\chi:=(\chi_{ij})$. Note that $\chi$ is simply the character table of the group. With this, we can see \cref{matrixeqn} as a matrix equation. 

As a preliminary step,  we obtain a \textbf{sum rule} for the $p_i$'s:
\begin{lemma}[Sum Rule for $p_i$'s]
    \begin{equation}
    \label{sumrule}
        \sum_i p_i \chi_{i}(e) = 1.
    \end{equation}
\end{lemma}
\begin{proof}
This follows from equation \ref{matrixeqn} applied to the conjugacy class $\{e\}$.
\end{proof}

We can solve for the $p_i$'s by inverting the matrix equation \ref{matrixeqn}. There is a trick to do this which is well known in the representation theory of groups: If one normalizes the $\chi_{ji}$'s, then one gets a unitary matrix. Namely,
\begin{equation}
    \hat{\chi}_{ji} = \chi_{ji} \cdot \frac{\sqrt{\# C_i}}{\sqrt{\# G}}
\end{equation}
defines a unitary matrix. Using unitarity, we can solve for the $p_i's$ by applying the adjoint of $\hat{\chi}$ to the matrix equation, and use well known properties of the character, to get that
\begin{equation}
    p_i(t) = \sum_j \frac{\sqrt{\# C_j}}{\sqrt{\# G}} e^{-t \,l(C_j)} (\hat{\chi}^{\dagger})_{ji} = \sum_j \frac{\# C_j} {\# G} e^{-t\, l(C_j)} \chi_{ij}^*.
\end{equation}

To answer question (2), we need to study the convexity of the Kraus-like decomposition. Since we already have a sum rule, we now need to find conditions which ensure that $p_i\geq 0$ for $1\leq i \leq n$.

Our following theorem gives the answer for question (2):
\begin{theorem}
\label{mainprop}
The character-induced Kraus-like decomposition of $P_t$ under the class function length $l$ is convex if and only if
\begin{equation}
    \label{semidef}
    p_i'(t=0) = -\sum_j \frac{\# C_j} {\# G} l(C_j) \chi_{ij}^* \geq 0
\end{equation}
for all $i \geq 2$.
\end{theorem}

We split the proof of Theorem \ref{mainprop} into two parts, necessity and sufficiency.

\begin{proposition}[Necessity]\label{necessity}
If the character-induced Kraus-like decomposition of $P_t$ is convex, then
\begin{equation}
   % \label{semidef}
    p_i'(t=0) = -\sum_j \frac{\# C_j} {\# G} l(C_j) \chi_{ij}^* \geq 0
\end{equation}
for all $i \geq 2$.
\end{proposition}
\begin{proof}
First observe that since $\chi_1$ is the character for the identity representation,  $\chi_1(C_i) = 1 = \chi_{1i}$ for all $i$, and so by the well known orthogonality of characters,
\begin{equation}
    p_i(t=0) = \langle \chi_i, \chi_1 \rangle = \delta_{i1}
\end{equation}
where $\langle f, g \rangle := \sum_{j} \frac{\# C_j} {\# G} f(C_j)^* g(C_j)$ for $f,g$ class functions on $G$.

Thus, if $p_i(t)$ is always positive, it is necessary that $p_i'(t=0) \geq 0$ for all $i\geq 2$.
\end{proof}

We will next show that this condition is actually sufficient for any finite group, and has a group-theoretical explanation. Before demonstrating the sufficiency of this condition in general, we show how one might go about it in a few specific cases. For the sake of our proofs, we state explicitly the nonnegative bounds for the length, even though it is explicit in the definition. We hope these examples highlight how quickly it becomes difficult to study the inequalities $p_i\geq 0$ due to the interdependency of the $p_i$. 

\subsubsection{Sufficiency of $p_i'(0)\geq 0$ for $S_3$}
	
The character table for $S_3$ is given by:

\begin{equation}
    \chi=\begin{pmatrix}
    1&1&1\\
    2&0&-1\\
    1&-1&1
    \end{pmatrix}
\end{equation}
with $\#C_1 = 1$, $\#C_2=3$, $\#C_3=2$. This yields 

\begin{equation*}
    \hat{\chi}=\frac{1}{\sqrt{6}}\begin{pmatrix}
    1&\sqrt{3}&\sqrt{2}\\
    2&0&-\sqrt{2}\\
    1&-\sqrt{3}&\sqrt{2}
    \end{pmatrix}
\end{equation*}

For convenience, denote $l(C_i)$ by $l_i$, and take $l_1=0$ so that the identity is of length $0$. 
This yields the following equations for the $p_i$:
\begin{align}
    p_1(t) &= \frac{1}{6} \left( 1+3 e^{-t l_2} +2 e^{-t l_3}\right)\\
    p_2(t) &= \frac{1}{6} \left(2-2e^{-tl_3} \right)\\
    p_3(t) &= \frac{1}{6} \left(1-3 e^{-t l_2} +2 e^{-t l_3}\right).
\end{align}

Note that $p_1$, $p_2$ are clearly non-negative for all $t\geq 0$. It is also automatic that $p_1'(0)\geq 0$ and $p_2'(0)\geq 0$. 

It is certainly necessary that $6p_3'(0)=3 l_2 - 2 l_3 \geq 0$. This will in fact turn out to be a sufficient condition.

\begin{proposition}
 In the set up for $S_3$ described above, all the $p_i$ are non-negative for all $t\geq 0$ if and only if $l_2\geq \frac{2}{3}l_3 \geq 0$.
\end{proposition}
\begin{proof}
The only if direction is clear.

For the if direction, as already noted, for any $t\geq 0$, the non-negativity of $p_1$ and $p_2$ is independent of the choice of $l_i$.

If $l_2 \geq \frac{2}{3} l_3$ we consider 2 cases

\begin{enumerate}
    \item Suppose $\frac{2}{3} l_3 \leq l_2 \leq l_3$. Then $$6p_3'(t) = 3 l_2 e^{-t l_2} -2 l_3 e^{-t l_3}\geq (3 l_2 - 2 l_3) e^{-t l_3} \geq 0.$$
    
    \noindent Thus, $p_3(0)=0$ and $p_3(t)$ is increasing, so $p_3(t)\geq 0$ for all $t\geq 0$.
    \item Suppose $l_2 \geq l_3$. Then $6p_3(t) \geq 1-3^{-t l_2} + 2 e^{-t l_2} = 1-e^{-t l_2} \geq 0$ for all $t\geq 0$.
\end{enumerate}
\end{proof}

As an example of what we have just shown, we will evaluate two natural notions of length on the group $S_3$. Let $|\cdot| = n-\text{\# of cycles}$. Then $l_1 = 3-3 = 0$, $l_2 = 3- 2=1$ and $l_3 = 3-1=2$. Clearly, $(l_1,l_2,l_3)=(0,1,2)$ violates the conditions of the above theorem and so it does not yield a probabilistic interpretation of the $p_i$. 

On the other hand, if $|\cdot| = \sqrt{n-\text{\# of cycles}}$, then $(l_1,l_2,l_3)=(0,1,\sqrt{2})$ and we do indeed obtain such a probabilistic interpretation.

\subsubsection{Sufficiency of $p_i'(0)\geq 0$ for $Q_8$}

The character table for $Q_8$ is 

\begin{equation}
    \chi=\begin{pmatrix}
    1&1&1&1&1\\
    1&1&1&-1&-1\\
    1&1&-1&1&-1\\
    1&1&-1&-1&1\\
    2&-2&0&0&0
    \end{pmatrix}
\end{equation}
where the conjugacy classes associated to the columns, in order, have sizes $1,1,2,2$ and $2$. This means that 
\begin{equation}
    \hat{\chi}=\frac{1}{2\sqrt{2}}\begin{pmatrix}
    1&1&\sqrt{2}&\sqrt{2}&\sqrt{2}\\
    1&1&\sqrt{2}&-\sqrt{2}&-\sqrt{2}\\
    1&1&-\sqrt{2}&\sqrt{2}&-\sqrt{2}\\
    1&1&-\sqrt{2}&-\sqrt{2}&\sqrt{2}\\
    2&-2&0&0&0.
    \end{pmatrix}
\end{equation}

Let $l_i=l(C_i)$, where we always take $l_1=0$ (i.e the identity element has length $0$). This yields the following expressions for the $p_i$: 
\begin{align}
    p_1=&\frac{1}{8}\left(1+e^{-tl_2}+2e^{-tl_3}+2e^{-tl_4}+2e^{-tl_5}\right)\\
    p_2=&\frac{1}{8}\left(1+e^{-tl_2}+2e^{-tl_3}-2e^{-tl_4}-2e^{-tl_5}\right)\\
    p_3=&\frac{1}{8}\left(1+e^{-tl_2}-2e^{-tl_3}+2e^{-tl_4}-2e^{-tl_5}\right)\\
    p_4=&\frac{1}{8}\left(1+e^{-tl_2}-2e^{-tl_3}-2e^{-tl_4}+2e^{-tl_5}\right)\\
    p_5=&\frac{1}{8}\left(2-2e^{-tl_2}\right)
\end{align}

It is clear that $p_1$ and $p_5$ are positive for all $t\geq 0$ and also that both $p_1'(0)$ and $p_5'(0)$ are non-negative. Note that when $t=0$, we have $p_2=p_3=p_4=0$. Let us focus momentarily on $p_2$. To make $p_2'(0)\geq 0$, we must have $-l_2-2l_3+2l_4+2l_5\geq 0$. The analogous computations for $p_3$ and $p_4$ show that $p_2'(0),p_3'(0)$ and $p_4'(0)$ are non-negative if and only if $l_2\leq \min\{2l_4+2l_5-2 l_3,2l_3+2l_5-2l_4,2l_3+2l_4-2l_5\}$. In fact, this condition turns out to be essentially sufficient.

\begin{proposition}
In the set up for $S_3$ described above, all the $p_i$ are non-negative for all $t\geq 0$ if and only if $0 \leq l_2\leq \min\{2l_4+2l_5-2 l_3,2l_3+2l_5-2l_4,2l_3+2l_4-2l_5\}$ and all the lengths are non-negative.
\end{proposition}
\begin{proof}

The only if direction is clear. For the if direction, as explained above, the non-negativity of $p_1$ and $p_5$ for $t\geq 0$ is independent of the choice of the $l_i$.

Note that $l_2$, $l_3$ and $l_4$ are symmetric in the equations for the $p_i$ in the sense that by relabeling the conjugacy classes, we can always assume WLOG that $l_3\leq l_4 \leq l_5$. In this case, $p_4\leq p_3$ and $p_4\leq p_2$, so it suffices to show that for any $t\geq 0$, $$8p_4(t)=1+e^{-tl_2} + 2(e^{-t l_5} - e^{-t l_3}-e^{-t l_4})\geq 0.$$

To this end, let $a=e^{-t l_3}$, $b=e^{-t l_4}$, $c=e^{-t l_5}$. By assumption, $e^{-tl_2}\geq \left(\frac{ab}{c}\right)^2$. Thus it suffices to show that $1+\left(\frac{ab}{c}\right)^2 + 2(c-a-b)\geq 0$ if $0\leq c\leq b\leq a \leq 1$.

First, we make a change of variable, setting $x = \frac cb\leq 1$. The desired inequality now reads
$2(a+b(1-x))\leq 1 + \frac{a^2}{x^2}$ for $0\leq b\leq a\leq 1$. Clearly, it suffices to check $b=a$, since the left-hand-side is monotonically increasing in $b$. So we only need to show that
$$
2a(2-x)\leq 1 + \frac{a^2}{x^2}.
$$

It is easy to see that the right-hand-side is at least $\frac{2a}{x}$ by the inequality of arithmetic and geometric means (AM-GM). Thus, our desired inequality is reduced to $4-2x\leq \frac{2}{x}$. This is true since $\frac{1}{x}+x\geq 2$, by AM-GM once more. 
\end{proof}

Extensions of these methods allow one to compute that for $S_4$, it is sufficient to have $p_j'(0)\geq 0$ for all $j\geq 2$ in order to guarantee the non-negativity of all the $p_i$'s. The computations for $S_4$ introduce new tools in addition to those used in the cases already presented, which may be useful for similar computations in larger groups. That being said, the proof for $S_4$ is significantly longer than the proof for the other groups, due to there being essentially three independent variables as opposed to two. The main additional technique involved in the proof for $S_4$ is a method to reduce the number of variables by Fourier-Motzkin elimination, which is an iterative approach. Due to the length of this computation, we relegate it to Appendix A.

Since the number of cases one needs to consider grows rapidly with the number of variable lengths involved, even with the algorithmic reduction method used for $S_4$, it is unlikely that the method is useful for any but low-dimensional groups. This leaves us looking for a more powerful, more general approach, which we take up next.

\subsubsection{Sufficiency of $p_i'(0)\geq 0$ for General Finite Groups}
Following the character-based method introduced in \cite{Lin}, we now prove that the  condition \begin{equation}
    \label{semidefinite}
    p_i'(t=0) = -\sum_j \frac{\# C_j} {\# G} l(C_j) \chi_{ij}^* \geq 0
\end{equation} for all $i\geq 2$ is actually sufficient to guarantee that $p_j(t)\geq 0$ for all $1\leq j\leq m$  and for all $t\geq 0$, where $m$ is the number of conjugacy classes of $G$, and equivalently the number of distinct irreducible representations of $G$. %The method introduced in \cite{Lin} will allow us to prove this fact.%only allows one to show that the linear \textit{definite} constraints are sufficient. We need a structural equivalence to extend the sufficiency of the definiteness constraints  to sufficiency of the semidefinite constraints, which we take up in the next section.} \Jon{I am confused by this paragraph. Maybe a more specific reference would help? like ``Theorem 5.\#"?}

 Let $\{\chi_r\}_{r=1}^m$ be the irreducible characters of $G$, and for each $r$, define $\sigma_r \lambda(g) = \chi_r(g) \lambda(g)$. Further assume that $l$ is a class function. Then, since characters span the class functions, $P_t$ admits a decomposition as $P_t = \sum_r p_r(t) \sigma_r$.

\begin{theorem}
	\label{maintheorem2}
	If there exists $\epsilon >0$ such that for all $1\leq r\leq m$, $p_r(t)\geq 0$ for all $0\leq t\leq \epsilon$, then for any $1\leq s\leq m$, $p_s(t) \geq 0$ for all $t\geq 0$.
\end{theorem}
\begin{proof}
		For any $t>0$, take $n$ large such that $t/n< \epsilon$. Then, $P_t = P_{n\left(\frac{t}{n}\right)} = \left(P_{\frac{t}{n}}\right)^n = \left( \sum_{i=1}^m p_i(t/n) \sigma_i\right)^n$ since $P_t$ is a semigroup. For $1\leq r\leq m$, let $\rho_r$ denote the irreducible representation with character $\chi_r$. The tensor product of irreducible representations of a finite group can be completely reduced, and the multiplicity of the irreducible representation $\rho_c$ in $\rho_a \otimes \rho_b$, $n_{ab}^c$, is always non-negative. By extension, we can write $\sigma_{a_1} \sigma_{a_2} \cdots \sigma_{a_n} = \sum_{i=1}^m n_{a_1 a_2 \ldots a_n}^{i} \sigma_i$, where $n_{a_1 a_2 \ldots a_n}^{i}$ is the multiplicity of the irreducible representation $\rho_i$ in $\rho_{a_1} \otimes \rho_{a_2} \otimes \cdots \otimes \rho_{a_n}$.  Thus,   
\begin{equation}
	P_t=\sum_{a_1,\cdots, a_n} \sum_{b} p_{a_1}(t/n) p_{a_2}(t/n) \cdots p_{a_n}(t/n) n_{a_1 a_2 \ldots a_n}^{b} \sigma_{b}.
\end{equation}
	Note that the coefficients in the above expression are all non-negative. Thus, for any $1\leq b\leq m$, we find that $p_b(t)$, the coefficient of $\sigma_b$, is nonnegative for all $t\geq 0$.
\end{proof}

\begin{cor}
\label{smallcor}
If there exists $\epsilon >0$ such that for all $1< r\leq m$, $p_r(t)\geq 0$ for all $t\leq \epsilon$, then for any $1\leq s\leq m$, $p_s(t) \geq 0$ for all $t\geq 0$. 
\end{cor}

\begin{proof}
Since $p_i(t=0) = \delta_{i1}$ and the $p_i$ are continuous, there is certainly a neighborhood of $0^{+}$ where $p_1(t)>0$. The conclusion then follows since all the hypotheses of Thm \ref{maintheorem2} are satisfied. 
\end{proof}

\begin{cor}
\label{smallercor}
If the $p_i'(t=0)$'s are {\color{black}positive} for all $i\geq 2$, then for any $1\leq s\leq m$, $p_s(t) \geq 0$ for all $t\geq 0$.
\end{cor}
\begin{proof}
The continuity of the $p_i$'s, combined with the positivity of the derivative, guarantees that there is a neighborhood of $0^{+}$ in which $p_i(t)>0$ for all $i\geq 2$. So Corollary \ref{smallcor} can be applied.
\end{proof}

{\color{black}Now we wish to strengthen Corollary \ref{smallercor} so that it suffices that all $p_i'(t=0)$'s are \textit{nonnegative} for all $i\geq 2$. To do so, it suffices to show that the set of lengths satisfying $p_i(t) \geq 0$ for all $t\geq 0$ is a closed set. This simply follows from the fact that the arbitrary intersection of closed sets is closed, applied to the set of lengths which satisfies $\{p_i(t_0)\geq 0\}$ for $t_0\in[0,\infty)$, in the Euclidean topology. We offer a different argument in the next section, which uses structural features from the condition of conditional negativity. This corollary, together with \Cref{necessity}, completes the proof of \Cref{mainprop}.

\section{Conditional Negativity Revisited}
In the previous section, it was shown that for the abelian finite groups, the notion of conditional negativity of a length function was equivalent with positing the existence of a convex Kraus-like decomposition. At the end of section 3.1, in particular, we used the need to understand this relationship on a more fundamental level to motivate our study of Kraus-like decompositions for general finite groups. In this section, we now give the characterization of the relationship between conditional negativity and the existence of a \textit{character-induced} convex Kraus-like decomposition in the full finite group case.

In the context of character-induced Kraus-like decompositions, the object of study is the decomposition of \textit{G-circulant} matrices \cite{Mor} $A=(A_{ij})$ with entries given by
\begin{equation}
	\label{matrixdef}
	A_{ij}= f(g_i g_j^{-1}) = \sum_{r \text{ an irrep}} p_r \chi_{r}(g_i g_j^{-1})
\end{equation}
where $f$ is a class function on $G$.
Such a decomposition exists since the irreducible characters form a basis of the set of class functions on $G$.

We wish to show the following theorem:

\begin{theorem}[Decomposition Theorem]\label{decompthm}
%The left-hand-side of equation \ref{matrixdef} makes up the entries of a positive-semi-definite matrix 
A G-circulant matrix is positive semidefinite 
if and only if the $p_r$'s arising in the decomposition are all nonnegative. 
\end{theorem}
% An equivalent statement is the following
% \begin{theorem}[Decomposition Theorem]
% Suppose $A$ is a G-circulant matrix. Then $A$ is positive-semidefinite if and only if the corresponding Kraus-like decomposition . See my proof below}
% \end{theorem}

By Schoenberg's theorem, if we can show this, then we obtain the following theorem:
\begin{theorem}
Suppose $l$ is a class function on $G$ satisfying $l(e)=0$. Then the corresponding character-induced Kraus-like decomposition is convex (i.e. the $p_r$'s are all nonnegative and satisfy a sum rule) if and only if $l$ is a conditionally negative-definite length. 
\end{theorem}

\begin{proof}[Proof of \Cref{decompthm}]
Define $A$ to be the matrix with entries $f(g_i g_j^{-1})$. We first prove the if direction. Observe that if the $p_r$'s are all non-negative, then $A$ is self-adjoint since
\begin{equation}
    A_{ij}=\sum_{r \text{ an irrep}} p_r \chi_{r}(g_i g_j^{-1})
\end{equation}
and 
\begin{equation}
    A_{ji}^* = \sum_{r \text{ an irrep}} p_r \chi_{r}(g_j g_i^{-1})^* = \sum_{r \text{ an irrep}} p_r \chi_{r}(g_j g_i^{-1})^*
\end{equation}
and we can write 
\begin{align}
    \chi_{r}(g_j g_i^{-1})^* &= (\sum_{n,m} r(g_j)_{n,m}r(g_i)^{-1}_{m,n})^* \\
    &=(\sum_{n,m} r(g_j)_{n,m}^*{r(g_i)^{-1}_{m,n}}^*) \\
    &= \sum_{m,n} r(g_j)^{\dagger}_{m,n} (r(g_i)^{-1})^{\dagger}_{n,m} \\
    &= \sum_{m,n} r(g_j^{-1})_{m,n} r(g_i)_{n,m}  \\
    &= \chi_r(g_i g_j^{-1})
\end{align}
where we have used the fact that the irreducible representations of finite groups are unitary.

Since $A$ is self-adjoint, it has a full spectrum. We want to show that the spectrum is nonnegative. It suffices to show that the matrices $(\chi_r(g_i g_j^{-1}))_{i,j}$, which by the above are self-adjoint, are in fact (up to normalization) orthogonal projections, in particular, that they satisfy 
\begin{equation}
	\label{desiredeqn}
	\sum_{j=1}^{|G|}\chi_r(g_i g_j^{-1}) \chi_s(g_j g_k^{-1}) = \delta_{r,s} \frac{|G|}{\chi_r(e)} 	\chi_r(g_i g_k^{-1}).
\end{equation}
 This would imply that the eigenvalues of $A$ are simply given by $p_r$'s up to positive rescaling (with additional multiplicities to account for the size of the subspace with the same eigenvalue). %{\color{blue}Then, since the $p_r$'s are, up to normalization, the spectrum of the matrix $A$, it follows that $A$ is positive semi-definite if and only if the $p_r$'s are nonnegative.  }

We prove this result using the idempotent method. Namely, it is known (see e.g., \cite{Janusz}) that within the group algebra of a finite group, one has the idempotents
\begin{equation}
	e_r = \frac{\chi_r(e)}{|G|} \sum_{g\in G} \chi_{r} (g^{-1}) \lambda(g)
\end{equation}
where $\lambda(g)$ is the left-regular representation of the finite group $G$. These idempotents satisfy
\begin{equation}
	e_r e_s = \delta_{r,s} e_r.
\end{equation}
Comparing coefficients of $\lambda(g)$ in $e_r e_r$ and $e_r$, one obtains that
\begin{equation}
	\label{maineqn}
	\frac{\chi_r(e)}{|G|} \sum_{h\in G} \chi_r(h g^{-1}) \chi_r(h^{-1}) = \chi_r(g^{-1})
\end{equation}
for all $g\in G$. Setting $g^{-1} = g_i g_k^{-1}$, and rewriting the dummy element $h$ as $h=g_j g_i^{-1}$ and summing over $g_j$, equation \ref{maineqn} becomes
\begin{equation}
	\frac{\chi_r(e)}{|G|} \sum_{g_j \in G} \chi_r(g_j g_k^{-1}) \chi_r(g_i g_j^{-1}) = \chi_r(g_i g_k^{-1})
\end{equation}
and so we have shown that equation \ref{desiredeqn} is true for $r=s$. When $r\neq s$, one has that $e_r e_s=0$, so the coefficient of each $\lambda(g)$ must vanish, yielding
\begin{equation}
		\frac{\chi_r(e)}{|G|} \sum_{g_j \in G} \chi_r(g_j g_k^{-1}) \chi_s(g_i g_j^{-1}) = 0.
\end{equation}
This completes the proof of equation  \ref{desiredeqn}.

For the only if direction, assume $A$ is positive semi-definite. Applying $A$ to an eigenvector $v$ of $(\chi_r(g_i g_j^{-1}))_{i,j}$ yields $p_r v$ up to a positive rescaling factor (since $(\chi_r(g_i g_j^{-1}))_{i,j}$ is a projection up to positive rescaling; we will fix the overall constant in the next section). So $p_r$ must be nonnegative. Since this must be true for any irrep $r$, the $p_r$'s must all be nonnegative.

\end{proof}

Bearing in mind that positive constant multiples of conditionally negative-definite lengths are conditionally negative-definite, we may summarize the theorems of this section as follows: 

\begin{theorem}
\label{bigtheorem}
Fix a group $G=\{g_1=e,\cdots,g_n\}$ and let $M$ be a $G$-circulant matrix such that $M_{1,1}=1$. Since the $\chi_r$ form a basis for class functions, there exists a decomposition $M_{ij}=\sum_{irreps}p_r\chi_r(g_ig_j^{-1})$. The following are equivalent:
\begin{itemize}
    \item $M$ is positive semi-definite.
    \item Each $p_r$ is non-negative.
    \item The length defined by $l(g_i)=-\ln (M_{i1})$ is conditionally negative-definite.
\end{itemize}
\end{theorem}

 % By Theorem \ref{bigtheorem},  it follows that if the length function $l$ is conditionally negative-definite, then so is $t l$.
 Thus, if we take $p_r$ to be the $p_r(t)$ \textit{induced} by the conditionally negative-definite length $l$ via our Kraus-like decomposition, then $p_r(t)\geq 0$ for all $t\geq 0$. Conversely, if $p_r(t) \geq 0$, then one obtains canonically a conditionally negative-definite length defined by $l(g_i) = - \frac{1}{t} \ln(M_{i1}(t))$. Hence, the set of class function lengths such that $P_t$ has a convex character-induced Kraus-like decomposition is precisely the set of conditionally negative-definite class function lengths. 

Based on this equivalence, we offer a different proof that the set of definite linear constraints we obtained to show the nonnegativity of all $p_i(t)$'s for all $t \geq 0$ in Corollary \ref{smallercor} can be improved to semidefinite linear constraints. 

The idea is to interpret that the length function $l$ as a point in a finite-dimensional space. Following Corollary \ref{smallercor}, one knows that the pre-image of the positive orthant $O$ under the map $\Phi(l)=(\Phi_2(l),\Phi_3(l), \cdots, \Phi_k(l))$, where $\Phi_i(l)\coloneqq-\sum_j \frac{\# C_j} {\# G} l(C_j) \chi_{ij}^* $ from $(l(C_2), \cdots, ,l(C_k))$ to $\mathbb{R}^{k-1}$, where $k$ is the number of conjugacy classes (recall that $l(C_1)=0$), is \textit{contained} in the set of conditionally negative-definite lengths. Note that $\Phi$ is a map to $k-1$ dimensional space and does not have a coordinate $\Phi_1(l)$, even though $\Phi_1(l)$ is defined.  %Furthermore, recall that the set of conditionally negative lengths is closed. Thus, we have the following proposition

We now prove two technical lemmas to support our proof that the definite constraints can be replaced by semi-definite constraints:

\begin{lemma}\label{condnegclosed}
Let $G$ be a group with $|G|=n$ and let $A\subset \mathbb{C}^{n}$ be the set of conditionally negative-definite lengths on $G$, where we identify a function $f:G\rightarrow\mathbb{C}$ with the vector $(f(g_1),\cdots, f(g_n))\in \mathbb{C}^n$. Then $A$ is closed. 
\end{lemma}

\begin{proof}
We show that $A^C$ is open. Let $f\in A^C$. Then there exists some $\{\alpha_i\}_{i=1}^n$ such that $\sum_{i=1}^n\alpha_i=0$ and 

\begin{equation*}
    \sum_{i,j=1}^n\alpha_i\overline{\alpha_j}f(g_i^{-1}g_j)>0.
\end{equation*}
This is an open condition, and so if we obtain a new length $f_\epsilon$ by changing $f$ by an $\epsilon$ amount in each coordinate, for $\epsilon$ sufficiently small, it will continue to hold with the same $\{\alpha_i\}_{i=1}^n$. Thus, $f_\epsilon\in A^C$ as well. This proves that $A^C$ is open and so also proves that $A$ is closed. 
\end{proof}

\begin{lemma}\label{phiinvert}
The map $\Phi$ is invertible.
\end{lemma}

% Just put \Phi's everywhere we see p's, and then replace e^{-tl} by -l
\begin{proof}

Note that in our proof, the condition \cref{sumrule}, which says  $\sum_{i=1}^k p_i\chi_i(e)=1$, translates to a linear condition on the $\Phi_i(l)$ which involves all the $\Phi_i(l)$ and in particular $\Phi_1(l)$. Thus, given $\Phi(l)=(\Phi_2(l),\ldots,\Phi_k(l))$, we can determine $\Phi_1(l)$.

Observe that \cref{semidefinite} gives an explicit expression for $\{\Phi_i(l)\}_{i=1}^k$. Note that it is immediately clear from character theory that ${\chi_{ij}^*}$ is invertible. It then follows that the map from $l$ to $( \Phi_1(l),\Phi_2(l),\ldots,\Phi_k(l))$ is invertible as well. Due to the addition of $\Phi_1$, note that this is not the map $\Phi$.

Putting this all together, it follows that given $x=(\Phi_2(l),\ldots,\Phi_k(l))$, we can determine $\Phi_1(l)$ and then uniquely recover $l=\Phi^{-1}(x)$.
\end{proof}

\begin{proposition}
Fix $l(C_1)=0$, and take $l(C_i)$ to be variable for $i=2,\ldots, k$. %Suppose the map $\Phi: \mathbb{R}^{k-1} \rightarrow \mathbb{R}^{k-1}$ defined by $\Phi(l)_i=-\sum_j \frac{\# C_j} {\# G} l(C_j) \chi_{ij}^* $ for $i=2,\ldots, k$ is invertible. 
If the $p_i'(t=0)$'s are {\color{black}nonnegative} for all $i\geq 2$, then for any $1\leq s\leq k$, $p_s(t) \geq 0$ for all $t\geq 0$.
\end{proposition}
\begin{proof}
Since $\Phi$ is invertible and linear, $\Phi^{-1}$ is continuous. Thus, $ \Phi^{-1}(\overline{O}) \subset  \overline{\Phi^{-1}(O)}$, where $O$ is the set of points with coordinates $x_i>0$ for all $i=1, 2,\ldots, k-1$. The latter is a subset of the set of conditionally negative-definite lengths (since this set is closed). Hence, any length whose image lies in $\bar{O}$ is conditionally negative-definite. Thus, we have strengthened the constraints from definite linear constraints (defined by $O$) to semidefinite linear constraints (defined by $\bar{O}$).
\end{proof}

\subsection{Values and Multiplicities of Eigenvalues}
We further show that the rank of $(\chi_r(g_i g_j^{-1}))_{i,j}$ is given by $\chi_r(e)^2$.  To prove this, first note that $\chi_r(g_i g_j^{-1}) = \chi_r(g_j g_i^{-1})^{*}$, so $(\chi_r(g_i g_j^{-1}))_{g_i,g_j}$ is Hermitian, and is fully diagonalizable. Furthermore, by rescaling equation \ref{desiredeqn}, we get that
\begin{equation}
	\label{rescaledeqn}
	\sum_{j=1}^{|G|} \left(\frac{\chi_r(e)}{|G|}\chi_r(g_i g_j^{-1})\right) \left(\frac{\chi_r(e)}{|G|}\chi_s(g_j g_k^{-1})\right) = \delta_{r,s} \frac{\chi_r(e)}{|G|} 	\chi_r(g_i g_k^{-1}),
\end{equation}
and so the $\frac{\chi_r(e)}{|G|}(\chi_r(g_i g_j^{-1}))_{g_i,g_j}$ is are orthogonal projection matrices, with eigenvalues $0$ or $1$. 
Thus, to compute its rank, we just need its trace, which is $|G|\cdot \frac{\chi_r(e)}{|G|}(\chi_r(e)) = (\chi_r(e))^2$. It follows that the multiplicity of the eigenvalue $1$ in the projection matrix corresponding to irrep $r$ is $(\chi_r(e))^2$.

Rewriting the decomposition of $A$ given by \cref{matrixdef} in terms of the projections, one has that
\begin{equation}
\label{matrixproj}
f(g_i g_j^{-1}) = \sum_{r \text{ an irrep}} \left(\frac{|G|}{\chi_r(e)}p_r\right) \left(\frac{\chi_r(e)}{|G|}\chi_{r}(g_i g_j^{-1})\right)
\end{equation}
and so the eigenvalues are given by $\left(\frac{|G|}{\chi_r(e)}p_r\right)$. Since these are orthogonal projections, we thus have shown that the multiplicity of each eigenvalue $\frac{|G|}{\chi_r(e)}p_r$ in the matrix given by \cref{matrixdef} is given by $(\chi_r(e))^2$.
 
\section{Conclusion}

In this article, we have presented a new way to approach semigroups acting on group algebras, namely, \textit{Kraus-like decompositions}. By modifying the action of possible operators which can be used in the sum decomposition of a semigroup, our approach allows us to introduce operator decompositions in problems where they were not readily available before. In particular, we are able to explore quantum channels induced by length functions in the context of group algebras. For length functions that are additionally class functions, we obtain several equivalent necessary and sufficient conditions on the length function for the semigroup $P_t$ to be a quantum channel for all time $t\geq 0$. 

Specifically, for \textit{character-induced} Kraus-like decompositions, we have proven that a set of semidefinite linear constraints is necessary and sufficient to guarantee the positivity of the coefficients appearing in the decomposition for all $t\geq 0$. We also proved that this same set of semidefinite linear constraints suffices to characterize the conditionally negative-definite lengths, a class of length functions that are of independent interest. Using Schoenberg's theorem, we further relate this to conditions on the complete positivity of semigroups $P_t$ induced by lengths which are class functions. These constraints follow from the fact that 
 for a semigroup $P_t$, the property of admitting a Kraus-like decomposition can be checked globally using a more local condition. Specifically, if one can show that $P_t$ admits a Kraus-like decomposition for $t>0$ sufficiently small, then the same can be concluded for all $t>0$.  

%We generalize full matrix algebra to group algebra

%Something about canonicalness of the multipliers, perhaps related to physical interpretation

%Include word quantum channel

%Convex Kraus like decompositions are "local" in that one only needs its for P_t with t small enough

The coefficients of our $p_i$'s are defined canonically. Moreover, they are nonnegative and satisfy a sum rule. Thus, the coefficients may have physical meaning (under appropriate rescaling) as \textit{probabilities}.  We plan to explore these physical connections in a follow-up paper.

\section*{Acknowledgments}
R.L. acknowledges the research support of ARO grant W911NF-20-1-0082 through the MURI project “Toward Mathematical Intelligence and Certifiable Automated Reasoning: From Theoretical Foundations to Experimental Realization.”  J.B. acknowledges the support of the Natural Sciences and Engineering Research Council of Canada (NSERC), [funding reference number 557353-2021]. J. B. a été financé par le Conseil de recherches en sciences naturelles et en génie du Canada (CRSNG), [numéro de référence 557353-2021].

The authors thank Radakrishnan Balu for helpful discussions at an intermediate stage of their work and helpful comments on the manuscript. The authors thank Arthur Jaffe for overall guidance, support, and feedback. They also thank Eric Carlen for bringing to their attention the complementary results of \cite{Wir}, which give structural characterizations of extensions of completely positive maps.

\appendix

\section{Proof of Stability Condition for $S_4$}
We have a direct proof that the a priori necessary conditions on the $p_i$ are sufficient to ensure the existence of a convex Kraus-like decomposition in the context of $S_4$ as well. We use methods that are similar to those presented above but more computationally involved due to the parameter space being of a larger dimension.

The character table for $S_4$ is 

\begin{equation*}
    \chi=\begin{pmatrix}
    1&1&1&1&1\\
    3&1&0&-1&-1\\
    2&0&-1&2&0\\
    3&-1&0&-1&1\\
    1&-1&1&1&-1
    \end{pmatrix}
\end{equation*}
and the sizes of the conjugacy classes are $(C_i)_{i=1,\ldots, 5} = (1,6,8,3,6)$.

Accordingly,
\begin{align*}
    p_1=&\frac{1}{24}\left(1+6e^{-tl_2}+8e^{-tl_3}+3e^{-tl_4}+6e^{-tl_5}\right)\\
    %p_2=&\frac{1}{24}\left(3e^{-tl_1}+6e^{-tl_2}+0-3e^{-tl_4}-6e^{-tl_5}\right)\\
    p_2=&\frac{1}{8}\left(1+2e^{-tl_2}-e^{-tl_4}-2e^{-tl_5}\right)\\
    %p_3=&\frac{1}{24}\left(2e^{-tl_1}+0-8e^{-tl_3}+6e^{-tl_4}+0\right)\\
    p_3=&\frac{1}{12}\left(1-4e^{-tl_3}+3e^{-tl_4}\right)\\
    %p_4=&\frac{1}{24}\left(3e^{-tl_1}-6e^{-tl_2}+0-3e^{-tl_4}+6e^{-tl_5}\right)\\
    p_4=&\frac{1}{8}\left(1-2e^{-tl_2}-e^{-tl_4}+2e^{-tl_5}\right)\\
    p_5=&\frac{1}{24}\left(1-6e^{-tl_2}+8e^{-tl_3}+3e^{-tl_4}-6e^{-tl_5}\right).
\end{align*}

Thus, at $t=0$, $p_1=1$ and all the others are equal to $0$. 

The constraint that the probabilities be nonnegative at time $\epsilon>0$ for $\epsilon$ arbitrarily small tells us that $p_i'(t=0)\geq 0$ for $2\leq i\leq 5$, In particular,
\begin{align*}
24p_2'(0)&=-6l_2 +3 l_4 + 6 l_5 \geq 0\\
24p_3'(0)&=8 l_3 -6 l_4\geq 0\\
24p_4'(0)&=6 l_2 + 3 l_4 - 6 l_5\geq 0\\
24p_5'(0)&=6 l_2 - 8 l_3 - 3 l_4 + 6 l_5\geq 0.
\end{align*}

We can clean this up a little bit by setting $a=e^{-t l_2}$, $b = e^{-t l_3}$, $c= e^{-t l_4}$, $d= e^{-t l_5}$.

\begin{theorem}
In the set up for $S_4$ described above, all the $p_i$ are non-negative for all $t\geq 0$ if and only if the lengths are all non-negative and the above necessary conditions hold. Expressed in terms of $a,b,c$ and $d$, this is equivalent to the following system of inequalities:

$$
0\leq a,b,c,d \leq 1.
$$
\begin{align*}
    a^{2} c^{-1} d^{-2} \geq 1 \\
    b^{-4} c^{3} \geq 1 \\
    a^{-2} c^{-1} d^{2} \geq 1 \\
    a^{-6} b^{8} c^{3} d^{-6} \geq 1.
\end{align*}.

\end{theorem}

\begin{proof}
We first express our goal, the non-negativity of the $p_i$, in terms of $a,b,c$ and $d$ as follows:
\begin{align*}
   1+2a-c-2d \geq 0 \\
   1-4b+3c \geq 0 \\
   1-2a-c+2d \geq 0 \\
   1-6a+8b+3c-6d \geq 0
\end{align*}
To prove these inequalities, we must consider a number of cases.

\textbf{The $a=d$ case}

We want to show that
\begin{align*}
    b^{-4} c^{3} \geq 1 \\
   a^{-12} b^{8} c^{3}  \geq 1
\end{align*}
implies
\begin{align*}
   1-4b+3c \geq 0 \\
   1-12a+8b+3c \geq 0,
\end{align*}
since the other inequalities become trivial in this case. The inequality $1-4b+3c\geq 0$ holds true since $1-4b+3c \geq 1-4b +3 b^{4/3}$, which is nonnegative on $[0,1]$ since its derivative is nonpositive and it evaluates to $0$ at $b=1$.

Note that $a \leq b^{2/3} c^{1/4}$, so $1-12 a+8b +3c \geq 1 - 12b^{2/3} c^{1/4} +8 b +3c$. We want to minimize this subject to $b^4 \leq c^3$, and show that the minimum is at least 0. If $b^4=c^3$, we obtain that the lower bound is given by $1-12b^{2/3}b^{1/3}+8b+3b^{4/3} = 1-4b+3 b^{4/3}$. This is the same situation as in the previous case and as such, is non-negative for $b\in[0,1]$. Taking a partial derivative with respect to $b$, we get $-8b^{-1/3}c^{1/4}+8\leq -8b^{-1/3}b^{1/3}+8= 0$, so $1 - 12b^{2/3} c^{1/4} +8 b +3c$ is non-negative at points $(b,c)$ where $b\in[0,c^{3/4}]$. This is precisely the range of values of $b$ allowed by the first inequality, so we are done.

\textbf{Relaxing the condition $a=d$:}
We now treat the full set of inequalities. We rewrite our assumptions in terms of $c$:
$$
c \leq \min(a^2 d^{-2}, a^{-2} d^{2})
$$
and
$$
c \geq \max(b^{4/3}, a^2 d^2 b^{-8/3} ).
$$

In particular, we have that 
$$
\max(b^{4/3}, a^2 d^2 b^{-8/3} ) \leq \min(a^2 d^{-2}, a^{-2} d^{2}),
$$
and so
\begin{align*}
b^{2/3} a&\leq d\leq b^{2/3}
\\
b^{2/3} d &\leq a \leq b^{2/3}
\end{align*}
are necessary conditions.

We first prove that $1+2a-c-2d \geq 0$. Note that everything is symmetric in $a$ and in $d$, so without loss of genereality, suppose $a\geq d$, then
$$
1+2a-c-2d \geq 1+2a - d^2/a^2-2d = 2(a-d) + 1-d^2/a^2 \geq 0.
$$

Next, we show $1+2d-c-2a \geq 0$. Continuing to assume that $a\geq d$, we note that 
$$
1+2d-c-2a \geq 1+2d - d^2/a^2 - 2a.
$$
We want to minimize $1+2d - d^2/a^2 - 2a$ over the region of permitted values of $a$ and $d$, so we start by fixing $b$ and $a$, and requiring $d$ to lie between $b^{2/3} a$ and $b^{2/3}$. (Note that we will be a bit lackadaisical with the constraint on $a$. This is OK since we just need a lower bound on the constraint region, which is guaranteed if we work with a region containing the constraint region.)

Taking a partial derivative with respect to $d$ yields $2-2d/a^2$ which changes sign at $d=a^2$, from positive to negative, so this is a maximum. We must evaluate this expression at the boundary of the allowed $d$ values as well. These are $d=b^{2/3}a$ and $d=a$ (since, by assumption, $d\leq a$). Evaluating at $d=b^{2/3}a$, we obtain $1-2 (1-b^{2/3}) a - b^{4/3} \geq 1-2(1-b^{2/3})b^{2/3}-b^{4/3} = 1-2b^{2/3} +b^{4/3} = (1-b^{2/3})^2\geq 0 $, where in the first inequality, we used $a\leq b^{2/3}$. For $d=a$, we get $1+2a-1-2a =0$. 

Next, we want to show that
$$
1-4b +3c \geq 0.
$$
Since $c\geq b^{4/3},$
we obtain $1-4b +3c \geq 1-4b+3b^{4/3} \geq 0$, as we showed earlier.

Finally, we want to show
$$
1-6(a+d)+8b+3c \geq 0.
$$
It is natural to split into two cases.

\textbf{Case 1: $b^2 \leq ad$:} In this case, we have $c \geq (ad)^2 b^{-8/3}$ as the stronger lower bound. Using this bound yields

$$1-6(a+d)+8b+3c \geq1-6(a+d) +8b + 3(ad)^2 b^{-8/3}.$$

We will minimize the sum $1-6(a+d) +8b + 3(ad)^2 b^{-8/3}$ for fixed $b$. We can still assume without loss of generality that $a\geq d$. Recall that $b^{2/3} a \leq d$. So we want to minimize
$1-6(a+d) +8b + 3(ad)^2 b^{-8/3}$ as a function of $a$ and $d$ subject to the two constraints
$$
b^{2/3} a \leq d \leq a \leq b^{2/3}
$$
and
$$
ad \geq b^2
$$

The shape of this domain is a three-sided region dependent on the value of $b$. In the plane with $a$ as its y-axis and $d$ as its x-axis, we have a region upper bounded by the line $a=b^{2/3}$, right-bounded by the line $d=a$, and southwest-bounded by the curve $ad=b^2$. The corners of the region are $(d=b^{4/3},a=b^{2/3})$, $(d=b^{2/3},a=b^{2/3})$, and $(d=b,a=b)$. 

For fixed $a$, we can differentiate with respect to $d$, obtaining $-6 + 6a^2 d b^{-8/3}$. Note that $a\geq b$ and $d\geq b^{4/3}$ on this region, so we have an upper bound on this derivative of $-6 + 6 b^2 b^{4/3} b^{-8/3} = -6 + 6 b^{2/3} \leq 0.$ Thus, the derivative is non-positive, and the function is non-increasing in $d$. As such, it will suffice to take $d$ lying on the boundary to the right, where $d=a$. 
We are thus left with considering $d=a$ in the expression $1-12a + 8b +3 a^4 b^{-8/3}$\footnote{Note that we have not considered this case yet. When we had $d=a$ earlier, we were directly considering the inequalities we wanted to prove, not the strengthened one we are working with here.}. Taking a derivative with respect to $a$, we obtain $-12 + 12 a^3 b^{-8/3}$, which equals $0$ when $a^3 b^{-8/3} = 1$, which is to say, when $a = b^{8/9}$. Note that $-12 + 12 a^3 b^{-8/3}<0$ for $a<b^{8/9}$ and $-12 + 12 a^3 b^{-8/3}>0$ for $a>b^{8/9}$, so we obtain a minimum at $a=b^{8/9}$ and it suffices to use this value going forward. Thus, we want to minimize $1-12 b^{8/9} + 8 b + 3 b^{32/9} b^{-8/3} = 1 - 9 b^{8/9} + 8 b.$ This has derivative $8-8 b^{-1/9} \leq 0$, so it suffices to verify $1 - 9 b^{8/9} + 8 b\geq 0$ at $b=1$, which certainly holds.

\textbf{Case 2: $b^2 \geq ad$:} Then $c \geq b^{4/3}$ is the meaningful lower bound on $c$. So we wish to minimize 
$$
1-6(a+d) +8b+ 3 b^{4/3}
$$
on the domain 
$$
b^{2/3} a \leq d \leq a \leq b^{3/2}
$$
and
$$
ad \leq b^{2}.
$$
This domain is also a three-sided region dependent on the value of $b$. It is bounded on the left by the line $d=b^{2/3} a$, on the right by the line $a=d$, on the northeast by the curve $ad=b^2$. The corners of the region are $(0,0)$, $(d=b^{4/3}, a=b^{2/3})$, and $(b,b)$.

The partial derivative of $1-6(a+d) +8b+ 3 b^{4/3}$ with respect to $d$ is $-6$, so due to the geometry of the domain boundary, this expression is minimized on the $d=a$ boundary if $a\leq b$ and on the $ad=b^2$ boundary if $a\geq b$. 

For $a\leq b$ and $d=a$, we have $1-6(a+d) +8b+ 3 b^{4/3}=1-12 a +8 b + 3 b^{4/3} \geq 1-4b+3b^{4/3}$. This is non-negative, as we showed earlier.

For $a\geq b$ and $ad=b^2$, we obtain $1-6(a+d) +8b+ 3 b^{4/3}=1-6( a+b^2/a) + 8 b + 3  b^{4/3} $. Taking a derivative with respect to $a$, one $-6(1-b^2/a^2)$. Thus, $1-6( a+b^2/a) + 8 b + 3  b^{4/3}$ has a maximum at $a=b$, and achieves its minimums at a boundary value of $a$. The boundary values of $a$ in the region we are considering are $a=b^{2/3}$ and $a=b$. If $a=b^{2/3}$, we get $1 + 8 b - 3 b^{4/3} - 6 b^{2/3}$, which has derivative $-\frac{4 (-1 + b^{1/3})^2}{b^{1/3}} \leq 0$. Thus, $1 + 8 b - 3 b^{4/3} - 6 b^{2/3}$ is minimized when $b=1$, yielding $0$. If $a=b$, we get $1-4b + 3  b^{4/3} $, which is non-negative, as shown earlier.
\end{proof}

\bibliographystyle{alpha}
\bibliography{bibliography}
\end{document}